\documentclass[a4paper,11pt]{article}
\pdfoutput=1 

\usepackage{jcappub} 

\usepackage[T1]{fontenc} 
\usepackage{graphicx}
\usepackage{epsfig}
\usepackage{rotate}
\usepackage{amsmath}
\usepackage{amssymb}
\usepackage{amsfonts}
\usepackage{bm}
\usepackage{tablefootnote}
\usepackage{enumerate}
\usepackage{afterpage}
\usepackage{natbib,hyperref}

\UseRawInputEncoding

\newcommand{\ud}{\mathrm{d}}

\def\be{\begin{equation}}
\def\ee{\end{equation}}
\def\bea{\begin{eqnarray}}
\def\eea{\end{eqnarray}}

\title{Constraining the growth rate on linear scales by combining SKAO and DESI surveys}

\author{Simthembile Dlamini${}^1 {}^,{}^a$, Sheean Jolicoeur$^1 {}^,{}^b$, Roy Maartens$^{1,2,3}{}^,{}^c$}
\affiliation{$^{1}$Department of Physics \& Astronomy, University of the Western Cape, Cape Town 7535, South Africa\\
$^{2}$Institute of Cosmology \& Gravitation, University of Portsmouth, Portsmouth PO1 3FX, United Kingdom\\
$^3$National Institute for Theoretical \& Computational Sciences (NITheCS), Cape Town 7535, South Africa }

\abstract{In the pursuit of understanding the large-scale structure of the Universe, the synergy between complementary cosmological surveys has proven to be a powerful tool. Using multiple tracers of the large-scale structure can significantly improve the constraints on cosmological parameters. We explore the potential of combining the Square Kilometre Array Observatory (SKAO) and the Dark Energy Spectroscopic Instrument (DESI) spectroscopic surveys to enhance precision on the growth rate of cosmic structures. We employ a multi-tracer Fisher analysis to estimate precision on the growth rate when using pairs of mock surveys that are based on SKAO and DESI specifications. The pairs are at both low and high redshifts. For SKA-MID, we use the HI galaxy and the HI intensity mapping samples. In order to avoid the complexities and uncertainties at small scales, we confine the analysis to scales where linear perturbations are reliable. The consequent loss of signal in each individual survey is mitigated by the gains from the multi-tracer. After marginalising over cosmological and nuisance parameters, we find a significant improvement in the precision on the growth rate.}

\emailAdd{$^{a}$simther4111@gmail.com} 
\emailAdd{$^{b}$jolicoeursheean@gmail.com}
\emailAdd{$^{c}$roy.maartens@gmail.com}

\begin{document}
\maketitle
\date{\today}
\flushbottom

\section{Introduction}

Einstein's theory of General Relativity (GR)  and  modified gravity  theories (see e.g. \cite{Clifton:2011jh, Koyama:2015vza, Langlois:2018dxi, Frusciante:2019xia}) prescribe the relation between peculiar velocities and the growth of large-scale structure. Peculiar velocities generate redshift-space distortions (RSD) in the power spectrum, which consequently provide a powerful probe for testing theories of gravity via the linear growth rate $f=-\ud \ln \ln D/\ud\ln(1+z)$, where $D(z)= \delta(z,\bm k) /\delta(0,\bm k)$ and $\delta$ is the matter density contrast. Here we assume that the  growth rate is scale-independent on linear scales. We confine our analysis to scales where linear perturbation theory is accurate, using a conservative $k_{\rm max}$. Although this leads to a significant loss in signal, it has the advantage that we can avoid the theoretical complexities and uncertainties involved in the modelling of small-scale RSD.

Precision measurements of RSD require the redshift accuracy of spectroscopic surveys. 
Currently, one of the best constraints on the growth index  is from the extended Baryon Oscillation Spectroscopic Survey (eBOSS)  survey, Data Release 14 Quasar \cite{Zhao:2018gvb}: 
\begin{align}\label{gamm}
\gamma \equiv \frac{\ln f(z)}{\ln\Omega_m(z)} = 0.580\pm 0.082\,.
\end{align}
This is consistent with the standard value $\gamma=0.55$, which is predicted by GR in the $\Lambda$CDM model. This value of $\gamma$  is also a good approximation for simple models of evolving dark energy, whose clustering is negligible \cite{Linder:2005in}. Statistically significant deviations from $\gamma=0.55$ could indicate either non-standard dark energy in GR or a breakdown in GR itself.
The next generation of multi-wavelength spectroscopic surveys (e.g. \cite{EUCLID:2011zbd,Chen:2012xu,Battye:2012fd,Newburgh:2016mwi,DESI:2016fyo,Santos:2017qgq,SKA:2018ckk,Sailer:2021yzm}) promises to deliver high-precision measurements of RSD, using complementary types of dark matter tracers. 

The effect of linear RSD on the power spectrum is degenerate with the amplitude of the matter power spectrum and with the linear clustering  bias. This degeneracy can be broken by using information
in the multipoles of the Fourier power spectrum (see e.g.  \cite{Bull:2015lja,Amendola:2016saw,Pourtsidou:2016dzn,Zhao:2018gvb,Castorina:2019zho}), or by using the angular power spectrum and including cross-bin correlations \cite{Fonseca:2019qek}. 
By combining information from different tracers, the multi-tracer technique \cite{Seljak:2008xr} can significantly improve  constraints on the growth rate \cite{McDonald:2008sh, White:2008jy, Blake:2013nif, Zhao:2020tis, Adams:2020dzw,Viljoen:2020efi}. 

We use Fourier power spectra 
in the flat-sky approximation. 
We perform a simple Fisher forecast on pairs of next-generation  spectroscopic surveys at low and at higher redshifts. The low-$z$ samples are similar to the Dark Energy Spectroscopic Instrument (DESI)
 Bright Galaxy Sample (BGS) \cite{DESI:2016fyo,Yahia-Cherif:2020knp,DESI:2022lza} and the Square Kilometer Array Observatory (SKAO) HI galaxies sample or the HI intensity mapping (IM) Band\,2 sample \cite{SKA:2018ckk,Berti:2022ilk}.
 For the higher-$z$ samples, we use samples similar to the DESI  Emission Line Galaxies (ELG) and  SKAO Band\,1 IM samples.
 
\section{Multi-tracer power spectra}
In redshift space, the positions of observed sources are made up of two parts. The first  is due to the background expansion of the universe, and the second is due to the peculiar velocities of the sources. Peculiar velocities are the result of the gravitational effect of local large-scale structure, and they induce shifts in the redshift-space positions of the sources. On large scales,  linear RSD produce an increase in clustering. For a given tracer $A$ of the dark matter distribution, the observed density contrast at linear order is 
\begin{equation}
\label{eq2.1}
\Delta_{A}(z,\bm{\hat{n}}) = \underbrace{b_{A}(z)\,\delta(z,\bm{\hat{n}})}_{\mbox{density}} - \underbrace{\frac{(1+z)}{H(z)}\,\hat{\bm n}\cdot \bm\nabla \big[\hat{\bm n}\cdot \bm v(z,\bm{\hat{n}})\big]}_{\mbox{linear RSD}}\;,
\end{equation}
where $b_{A}$ is the linear bias, 
$\bm v$ is the peculiar velocity, and $\bm{\hat{n}}$ is the unit vector in the line of sight direction of the observer.
In the flat-sky approximation (fixed $\bm{\hat{n}}$), the Fourier transform of \eqref{eq2.1} gives
\begin{equation}
\label{eq2.2}
\Delta_{A}(z,\bm{k}) = \Big[b_{A}(z) + f(z)\mu^{2}\Big] \delta (z,\bm{k})
\quad\mbox{where} \quad \mu = \bm{\hat{k}} \cdot \bm{\hat{n}}\;.
\end{equation}
Here we  used the first-order continuity equation 
\begin{equation}
\label{eq2.3}
\bm\nabla\cdot \bm v = - \frac{H}{(1+z)}\, f\, \delta \;.
\end{equation}
The tree-level Fourier power spectra are then defined by 
\begin{equation}
\label{eq2.5}
\big \langle \Delta_{A}(z,\bm{k})\,\Delta_{B}(z,\bm{k}') \big \rangle = (2\pi)^{3}\,P_{A B}(z,\bm{k})\,\delta^{\mathrm{D}}(\bm{k}+\bm{k}')\;.
\end{equation}
By \eqref{eq2.2},  
\begin{equation}
\label{eq2.6}
P_{A B}(z,\bm{k}) = P_{A B}(z,k,\mu) = \Big[b_{A}(z) + f(z)\mu^{2}\Big]\Big[b_{B}(z) + f(z)\mu^{2}\Big] P(z,k)\;, 
\end{equation}
where $P$ is the linear matter power spectrum (computed from CLASS \cite{Blas:2011rf}). We can split it into a shape function $\mathcal{P}$ and an amplitude parameter $\sigma_{8,0}$ as:
\begin{equation}
\label{eq2.7}
P(z,\bm k) = \sigma_{8,0}^{2}\,\mathcal{P}(z,\bm k)\;.
\end{equation}

{Note that in general there is a scale-dependent cross-correlation coefficient, $0<r\leq1$, that multiplies the $P_{AB}$ in \eqref{eq2.6} \cite{Wolz:2015ckn,Anderson:2017ert}. On the large, linear scales that we consider, it is expected that $r$ can be taken to be 1 (e.g. \cite{Rubiola:2021afc}).}

\subsection{Sample specifications}
We consider mock samples similar to the following spectroscopic samples:
\begin{itemize}
\item galaxies: DESI BGS and ELG \cite{DESI:2016fyo,Yahia-Cherif:2020knp} and SKAO Band\,2 HI galaxies \cite{SKA:2018ckk}.
\item intensity mapping: SKAO HI IM Band\,1,2 \cite{SKA:2018ckk}.
\end{itemize}
\autoref{table1}, based on \cite{DESI:2016fyo, SKA:2018ckk}, shows the sky and redshift coverage of the individual and overlapping samples, together with the survey time for the HI samples. For the overlap sky areas, we assume nominal values.
\begin{table}[!ht] 
\centering 
\caption{Sky area and redshift range of samples, and survey time for HI samples.} \label{table1} 
\vspace*{0.2cm}
\begin{tabular}{|l|l|c|c|c|} 
\hline 
Survey~~&~~Sample~~  & ~~$\Omega_{\mathrm{sky}}$~~ &~~$t_{\mathrm{tot}}$~~ & ~~redshift~~ \\ 
&  & $\big[10^{3}\,\mathrm{deg}^{2}\big]$ & $\big[10^{3}\,\mathrm{hr}\big]$ & ~~range~~ \\\hline\hline 
$g$~(DESI-like) & ELG & 14 & - & 0.60--1.70  \\
                & BGS & 14 & - & 0.00--0.50 \\
$g$~(SKAO-like) &HI gal. & 5 & 10 & 0.00--0.50 \\ 
\hline \hline
$H$~(SKAO-like)                &IM Band\,1 & 20 & 10 & 0.35--3.05\\
                &IM Band\,2 & 20 & 10 & 0.10--0.58\\\hline \hline
$g_1\times g_2$ & BGS $\times$ HI gal.   & 5  & 5 & 0.00--0.50 \\
$g\times H$ &ELG $\times$ IM Band\,1 & 10 & 5 & 0.60-1.70 \\                
 $g\times H$           &BGS $\times$ IM Band\,2 & 10 & 5 & 0.10--0.50 \\\hline
\end{tabular}
\end{table}

For the linear clustering biases $b_{A}$, we use one-parameter models
where the redshift evolution  is assumed known,
as suggested by \cite{Agarwal:2020lov}.  For the DESI-like samples we use \cite{Jelic-Cizmek:2020pkh}:
\begin{equation}
 \label{eq3.1}
b_{g}(z) = \frac{b_{g0}}{D(z)} \quad \text{with fiducial value}~ b_{g0}=1.34~\mbox{(BGS)}\quad \mbox{and}\quad 0.84~\mbox{(ELG)} \,.
\end{equation}
For the SKAO-like HI galaxy sample, we use \cite{SKA:2018ckk}:
\begin{equation}
 \label{eq3.1a}
b_{g}(z) = {b_{g0}}\big(1 + 0.880\,z - 0.739\,z^{2}\big) \quad \text{with fiducial value}~ b_{g0}=0.625 \,.
\end{equation}
For HI IM, we use a fit based on simulations \cite{Villaescusa-Navarro:2018vsg}:
\begin{equation}
\label{eq3.2}
b_{H}(z) =
b_{H0}\big(1 + 0.693\,z - 0.046\,z^{2}\big)  \quad \text{with fiducial value}~ b_{H0}= 
0.842\,.
\end{equation}
The background brightness temperature of HI IM is modelled via the fit given in \cite{Santos:2017qgq}:
\begin{equation}
\label{eq3.4}
\bar{T}_{H}(z) = 0.0559 +0.2324\,z -0.0241\, z^{2} ~~ \mathrm{mK}\,.
\end{equation}

\subsection{Noise}
\begin{figure}[! ht]
\centering
\includegraphics[width=7.0cm]{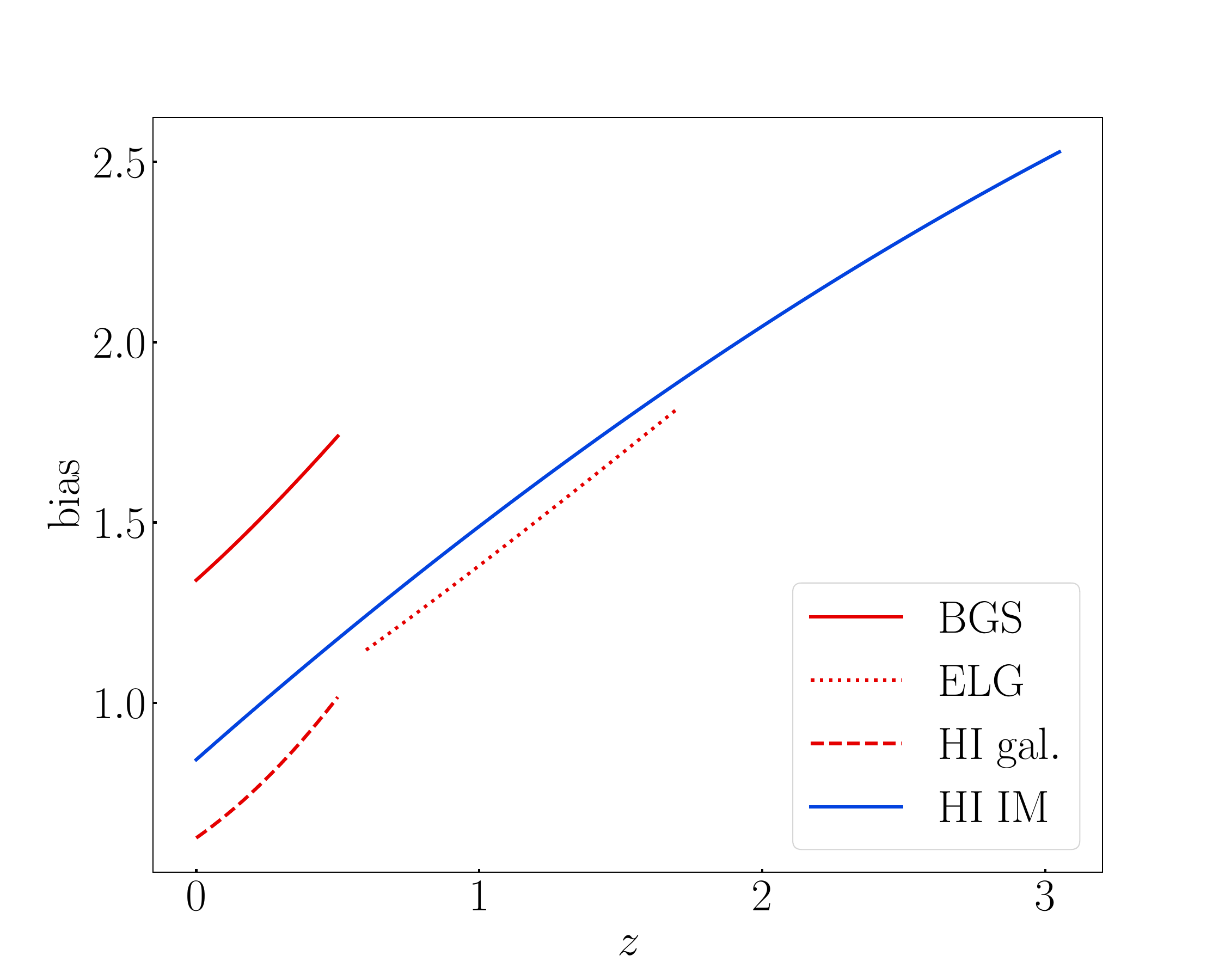} 
\includegraphics[width=7.7cm]{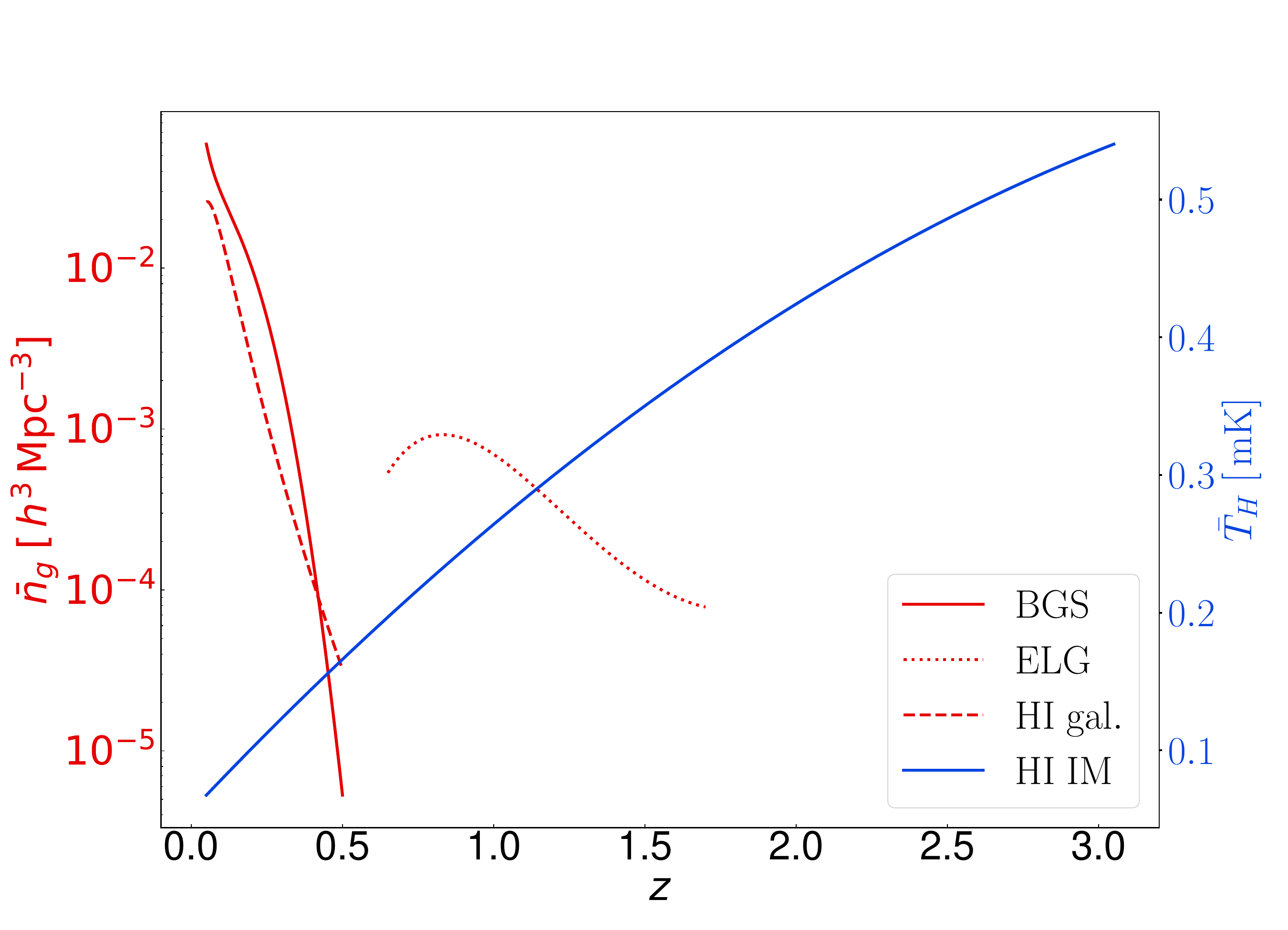}
\caption{\emph{Left:} Fiducial clustering bias for  galaxy (red) and intensity mapping  (blue) samples. \emph{Right:} Comoving number density for galaxy surveys ({red}, left $y$-axis) and brightness temperature for IM surveys ({blue}, right $y$-axis)} 
\label{fig1}
\end{figure}

For galaxy surveys, the noise that affects the auto-power spectrum measurement is the shot noise (assumed to be Poissonian):
\begin{equation}
\label{eq3.5}
P^{\rm shot}_{gg}(z) = \frac{1}{\bar n_{g}(z)}\;,
\end{equation}
where $\bar n_{g}$ is the comoving background number density. 
The total signal for the galaxy auto-power spectra is
\begin{equation}
\label{eq3.6}
\tilde{P}_{gg}(z,k,\mu) = P_{gg}(z,k,\mu) + P^{\rm shot}_{gg}(z)\;.
\end{equation}
\autoref{fig1} shows the fiducial clustering biases and number density and brightness temperature for all the samples.

There is shot noise in HI IM surveys --  but on the linear scales that we consider, this  shot noise is much smaller than the thermal noise (see below) and can be safely neglected \cite{Castorina:2016bfm, Villaescusa-Navarro:2018vsg}. 

For the cross-power spectra, the cross shot-noise may be neglected if the overlap of halos hosting the two samples is negligible. This is shown to be the case for BGS\,$\times$\,IM in \cite{Viljoen:2020efi} (see also \cite{Casas:2022vik}). We assume that it is a reasonable approximation in the cases ELG\,$\times$\,IM and BGS\,$\times$\,HI galaxies, so that $P_{gH}^{\rm shot}\approx 0\approx P_{gg'}^{\rm shot}$. (Note that we do not consider the multi-tracer case HI galaxies\,$\times$\,HI IM.)

The thermal noise in HI IM  depends on the sky temperature in the radio band, the survey specifications and the array configuration (single-dish or interferometer). For the single-dish  mode of SKAO-like IM surveys, the thermal noise power spectrum is  \cite{Bull:2014rha,Alonso:2017dgh,Jolicoeur:2020eup}:
\begin{equation}
\label{eq3.10}
P_{H\!H}^{\mathrm{therm}}(z) = \frac{\Omega_{\mathrm{sky}}}{2\nu_{21} t_{\mathrm{tot}}}\,\frac{(1+z)^2 r(z)^{2} }{H(z)} \,\left[\frac{T_{\mathrm{sys}}(z)}{\bar{T}_{H}(z)}\right]^2\, \frac{1}{N_{\rm d}}\,, 
\end{equation}
where $\nu_{21}=1420\,\mathrm{MHz}$ is the rest-frame frequency of the 21\,cm emission,  $t_{\mathrm{tot}}$ is the total observing time, and the number of dishes is $N_{\mathrm{d}}=197$ (with dish diameter  $D_{\mathrm{d}}=15\,$m). The system temperature is modelled as \cite{Ansari:2018ury}:
\begin{equation}
\label{eq3.11}
T_{\mathrm{sys}}(z) = T_{\rm d}(z)+T_{\rm sky}(z) =T_{\rm d}(z) + 2.7 + 25\bigg[\frac{400\,\mathrm{MHz}}{\nu_{21}} (1+z)\bigg]^{2.75} ~ \mathrm{K}, 
\end{equation} 
where $T_{\rm d}$ is the dish receiver temperature given in \cite{Viljoen:2020efi}. The total signal is then 
\begin{equation}
\label{eq3.12}
\tilde{P}_{H\!H}(z,k,\mu) = 
P_{H\!H}(z,k,\mu) + P_{H\!H}^{\mathrm{therm}}(z)\,.
\end{equation}
 
\subsection{Intensity mapping beam and foregrounds}
\label{SS4.2}
HI IM surveys in single-dish mode have poor angular resolution, which results in power loss on small transverse scales, i.e. for large $k_\perp=(1-\mu^2)^{1/2}k$. This effect is typically modelled by a Gaussian beam factor \cite{Bull:2014rha}:
\begin{equation}
\label{eq3.14}
{\cal D}_{\rm beam}(z,k,\mu)=\exp\left[-\frac{(1-\mu^2)k^2 r(z)^2\theta_{\rm b}(z)^2}{16\ln 2} \right] 
\quad\mbox{with}\quad \theta_{\rm b}(z) = 1.22\,\frac{\lambda_{21}(1+z)}{D_{\rm d}}\,.
\end{equation}

HI IM surveys are also contaminated by foregrounds much larger than the HI signal. 
Since these foregrounds are spectrally smooth, they can be separated from the non-smooth signal on small to medium scales. However, on very large radial scales, i.e. for small $k_\|=\mu k$, the signal becomes smoother and therefore, the separation fails.
A comprehensive treatment requires simulations of foreground cleaning of the HI signal (e.g. \cite{Cunnington:2020wdu,Spinelli:2021emp}). 
For a simplified Fisher forecast, we can instead use a foreground avoidance approach by excising the regions of Fourier space where the foregrounds are significant.  This means removing large radial scales, which can be modelled by the foreground-avoidance  factor: 
\begin{equation}
\label{eq3.15}
\mathcal{D}_{\mathrm{fg}}(k, \mu) = \Theta\left(\left|k_{\|}\right|-k_{\mathrm{fg}}\right)= \begin{cases}1, & \left|k_{\|}\right|>k_{\mathrm{fg}} \\ 0, & \left|k_{\|}\right| \leq k_{\mathrm{fg}}\end{cases}
\end{equation}
where $\Theta$ is the Heaviside step function. We assume the cut is made at a minimum value of 
\begin{equation}
\label{eq3.16}
k_{\mathrm{fg}} = 0.01\, h\, \mathrm{Mpc}^{-1}\,.
\end{equation}

In summary, the HI IM density contrast is modified by beam and foreground effects as follows:
\begin{align}
 \Delta_H(z,k,\mu) ~~\to~~ \mathcal{D}_{\rm beam}(z,k,\mu) \, \mathcal{D}_{\mathrm{fg}}(k,\mu)\, \Delta_H(z,k,\mu)\,.
\end{align}

\section{Multi-tracer Fisher Analysis}
\label{S4.1}
The Fisher matrix in each redshift bin for the combination of two dark matter tracers is \cite{Barreira:2020ekm,Karagiannis:2023lsj}
\begin{equation}
\label{eq4.1}
F_{\alpha \beta}^{\bm P} = 
\sum_{\mu=-1}^{+1}
\,\sum_{k=k_{\mathrm{min}}}^{k_{\mathrm{max}}}\,\partial_{\alpha}\,\bm{P} \cdot \mathrm{Cov}(\bm{P},\bm P)^{-1} \cdot \partial_{\beta}\,\bm{P}^{\mathrm{T}}\;,
\end{equation}
where $\partial_{\alpha} = \partial\,/\,\partial \vartheta_{\alpha}$, with $\vartheta_{\alpha}$ the parameters, and $\bm P$ is the data vector of the power spectra:
\begin{equation}
\label{eq4.2}
\bm{P} = \big( P_{g g}\,,\,  {P}_{g H}\,,\,  {P}_{H\!H}\big) \quad\mbox{or}~~\big( P_{g g}\,,\,  {P}_{g g'}\,,\,  {P}_{g'g'}\big)\;.
\end{equation}

Note that the sum over $\mu$ incorporates the foreground avoidance via the Heaviside factor \eqref{eq3.15} in $P_{HA}$. Also note that $\bm P$ contains no noise terms -- these appear in the covariance below. The reason is that noise does not depend on the cosmological parameters. Although the thermal noise in HI IM depends on $H$, this arises from mapping the Gaussian pixel noise term to Fourier space. 
%

The multi-tracer covariance includes the shot and thermal noises, and is given by \cite{White:2008jy,Zhao:2020tis, Barreira:2020ekm,Karagiannis:2023lsj}: 
\begin{equation}
\label{eq4.5}
\mathrm{Cov}(\bm{P},\bm P) =  \frac{k_{\rm f}^3}{2\pi k^{2} \Delta k}\,\frac{2}{\Delta \mu}\,
\begin{pmatrix}
\tilde P_{gg}^2 & & \tilde P_{gg}\tilde P_{gH} & & \tilde P_{gH}^2 \\ 
& & & & \\
\tilde P_{gg}\tilde P_{gH} & & \frac{1}{2}\big[\tilde P_{gg}\tilde P_{H\!H}+ \tilde P_{gH}^2 \big] & & \tilde P_{H\!H}\tilde P_{gH} \\
& & & & \\
\tilde P_{gH}^2 & & \tilde P_{H\!H}\tilde P_{gH} & & \tilde P_{H\!H}^2 
\end{pmatrix}\;,
\end{equation}
and similarly for the case $g\times g'$.
Here $\Delta k$ and $\Delta \mu$ are bin-widths and the fundamental mode $k_{\mathrm{f}}$, corresponding to the longest wavelength, is determined  by the comoving survey volume of the redshift bin centred at $z$:
\begin{equation}
\label{eq4.6}
V(z) = \frac{\Omega_{\mathrm{sky}}}{3} \bigg[r\Big(z+\frac{\Delta z}{2}\Big)^{3} - r\Big(z-\frac{\Delta z}{2}\Big)^{3} \bigg]=
\bigg[\frac{2\pi}{k_{\mathrm{f}}(z)} \bigg]^3
\;.
\end{equation}

We choose the bin-widths following \cite{Karagiannis:2018jdt,Yankelevich:2018uaz,Maartens:2019yhx}:
\begin{equation}
\label{eq4.7}
\Delta z = 0.1, ~~\Delta \mu = 0.04, ~~ \Delta k = k_{\mathrm{f}}\;.
\end{equation}

In order to exclude the small length scales that are beyond the validity of linear perturbation theory, we impose a conservative maximum wavenumber of $.08h/$Mpc at $z=0$, with a redshift evolution as proposed in \cite{Smith:2002dz}:
\begin{equation}
\label{eq4.8}
k_{\mathrm{max}}(z) = 0.08\,(1+z)^{2/(2+n_{s})} \;h\,\mathrm{Mpc}^{-1}\;.
\end{equation}

The largest length scale that can be measured in galaxy surveys corresponds to the smallest wavenumber, given by 
\begin{equation}
\label{eq4.9}
k_{\mathrm{min}} = k_{\mathrm{f}}\,.
\end{equation}
For HI IM surveys, 
$k_{\mathrm{min}} = \text{max}\big(k_{\mathrm{f}}, k_{\mathrm{fg}}\big)$.\\

The multi-tracer Fisher matrix applies for a perfectly overlapping region in both redshift range and sky area for the two tracers. If the samples differ in redshift and sky area, then we can add the independent non-overlapping Fisher matrix information of the individual surveys. The full Fisher matrix, denoted by $g\otimes H$, is \cite{Viljoen:2020efi,Jolicoeur:2023tcu}
\begin{equation}
\label{eq4.11}
F_{\alpha \beta}^{g \otimes H} = 
F_{\alpha \beta}^{\bm P}(\text{overlap})+
F_{\alpha \beta}^{g}(\text{non-overlap}) + F_{\alpha \beta}^{H}(\text{non-overlap})\;,
\end{equation}
and similarly for the case $g\otimes g'$.

For the cosmological parameters, we choose $\sigma_{8,0}$, $\Omega_{b,0}$, $\Omega_{c,0}$, $n_s$ and $h$, since we are focusing on  constraining the growth rate index $\gamma$, which should be minimally affected by the remaining $\Lambda$CDM parameters on linear scales. We fix these remaining cosmological parameters to their Planck 2018 best-fit values \cite{Planck:2018vyg}. We therefore consider the following set of cosmological parameters together with the two nuisance bias parameters:
\begin{equation}
\label{eq4.3}
\vartheta_{\alpha} = \big(\gamma, \, \sigma_{8,0}, \,n_{s}, \,h, \,\Omega_{b0},\, \Omega_{c0}\,;\,b_{A0}\big)\quad \mbox{where}~~
A= \mbox{BGS, ELG, HIg, IM1, IM2} \;.
\end{equation}

The marginalised errors are then obtained as
\begin{equation}
\label{eq4.4}
\sigma\big(\vartheta_{\alpha}\big) = \Big[\big(F^{-1}\big)_{\alpha \alpha}\Big]^{1/2}\;.
\end{equation}

We compute numerically the Fisher derivatives with respect to $n_{s}$, $h$, $\Omega_{b0}$ and $\Omega_{c0}$, using the 5-point stencil approximation with selected step sizes, shown in \autoref{tab2}. The derivatives are stable for $0.0003 \le \mbox{step size} \le 0.1$ \cite{Yahia-Cherif:2020knp}. The derivatives with respect to $\gamma$, $\sigma_{8,0}$ and $b_{A0}$ are computed analytically, with for example
\begin{align}
\frac{1}{P_{AA}}\, \partial_\gamma P_{AA}=
\frac{2\mu^2\,\Omega_m^\gamma\,\ln \Omega_m}{b_A+\mu^2\,\Omega_m^\gamma }\,.
\end{align}

\begin{table}[!h]
\centering
\caption{ {Numerical step sizes for Fisher derivatives.}}  
\label{tab2}
\vspace*{0.2cm}
\begin{tabular}{|lcccc|} \hline
Parameters & $n_{s}$ & $h$ & $\Omega_{b0}$ & $\Omega_{c0}$  \\ 
Optimal step size & 0.1 & 0.01 & 0.01 & 0.02  \\
\hline
\end{tabular}
\end{table}

 \section{Results}
\begin{figure}[! ht]
\centering
\includegraphics[width=13.5cm]{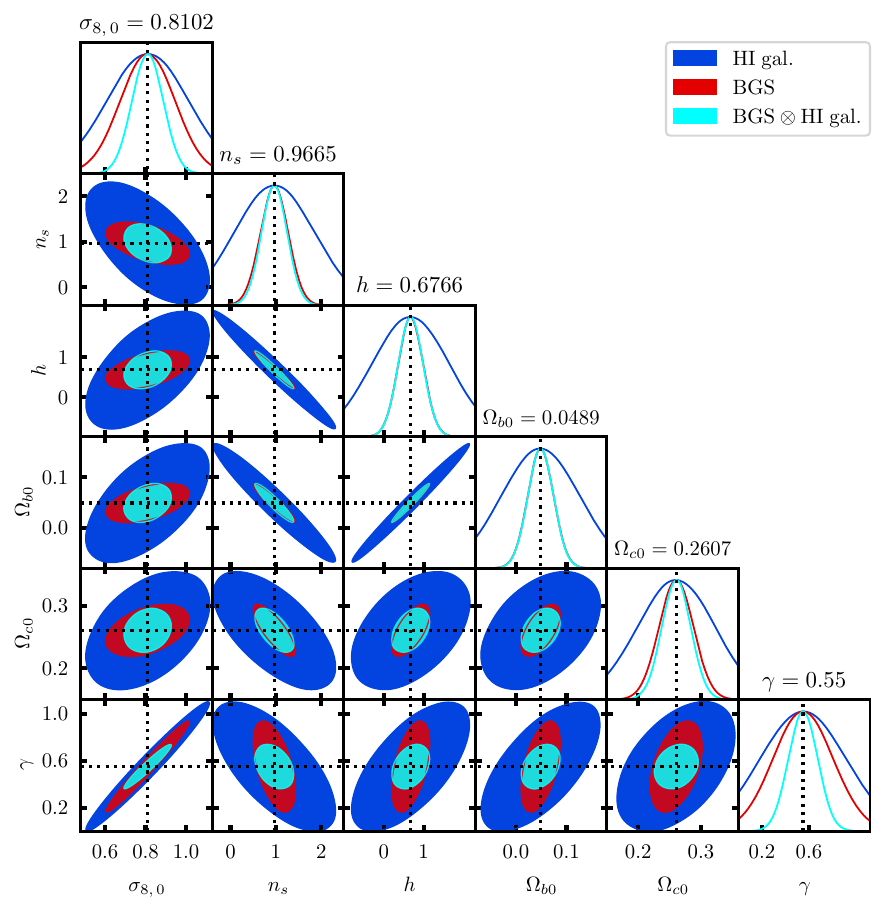}  
\caption{Marginal 1$\sigma$ contours for $\gamma$ and cosmological parameters, from the BGS and HI galaxy combination.} 
\label{fig3}
\end{figure}

\begin{figure}[! ht]
\centering
\includegraphics[width=13.5cm]{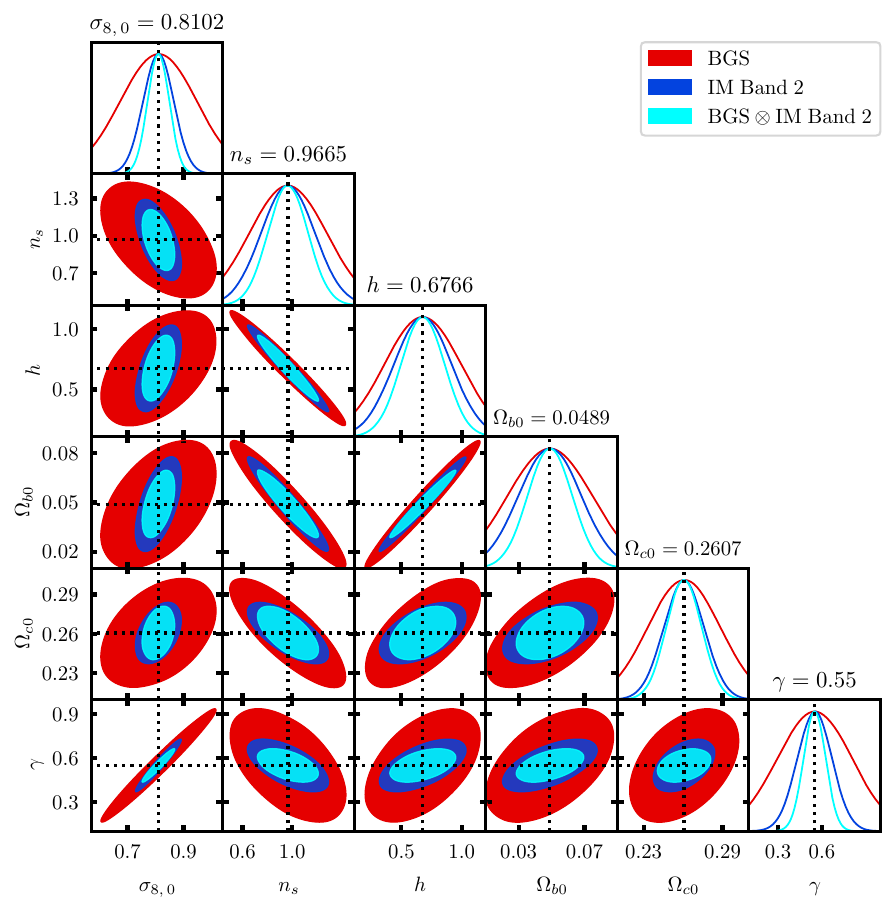} 
\caption{As in \autoref{fig3}, from the BGS and HI IM Band\,2 combination.}
 \label{fig5}
\end{figure}

\begin{figure}[! ht]
\centering
\includegraphics[width=13.5cm]{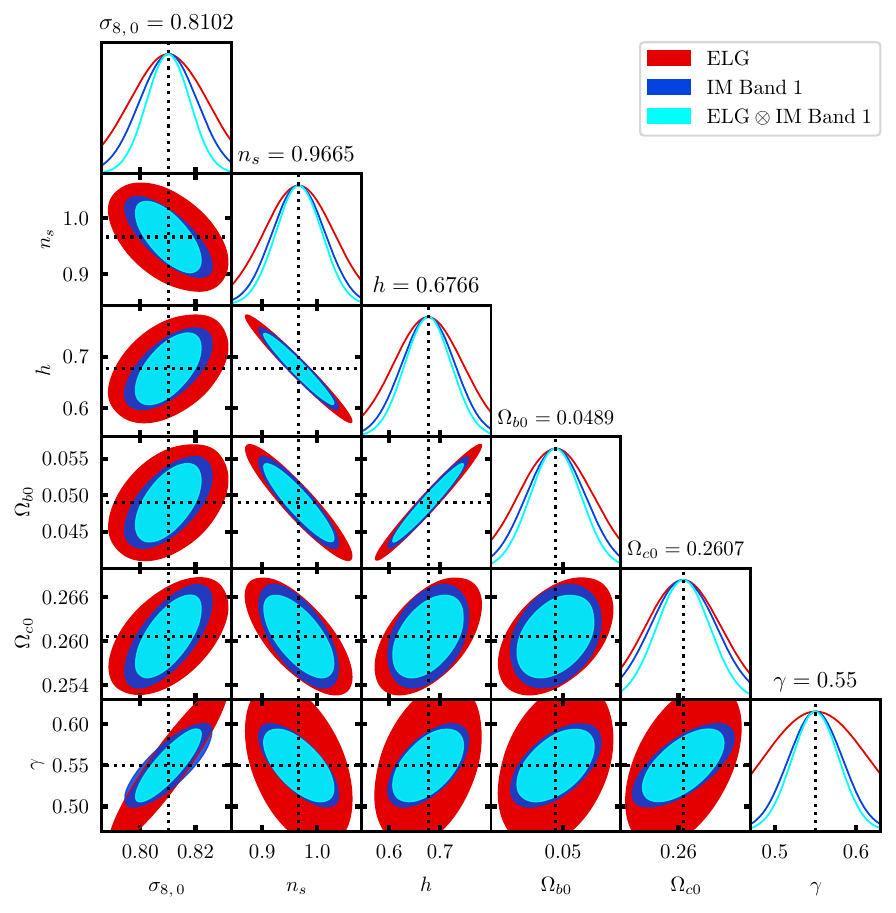} 
\caption{As in \autoref{fig3}, from the ELG and HI IM Band\,1 combination.}
\label{fig4}
\end{figure}

\autoref{fig3}--\autoref{fig4} show the 1$\sigma$ error contours for the parameter $\gamma$ and the cosmological parameters, after marginalising over the 2 bias nuisance parameters in \eqref{eq4.3}. There are significant degeneracies, which the multi-tracer partly alleviates, allowing for improved precision on the cosmological parameters. The improvement is small, unlike the case of the growth index, which shows significant improvement. This is not surprising, since the multi-tracer removes cosmic variance from the effective bias, i.e. the clustering bias plus the RSD contribution, as shown in \cite{McDonald:2008sh}.
All multi-tracer pairs show a significant reduction in errors on $\gamma$ compared to the best single tracer.  The improvements are shown in the fractional errors listed in \autoref{tab3}. We note that these multi-tracer errors are obtained using only linear scales.

For the BGS and HI galaxy combination in \autoref{fig3}, the multi-tracer fractional error on $\gamma$ is less than half of the BGS value. We note that our constraint on $\gamma$ for the single-tracer BGS is weaker than that in \cite{Viljoen:2020efi}. The reason is that \cite{Viljoen:2020efi} uses the angular power spectra with a large number of very thin redshift bins (width 0.01) and considers all possible cross-bin correlations. By contrast, our standard Fourier analysis only uses 5 redshift bins of width 0.1, and does not include cross-correlations between different redshift bins. The single-tracer HI galaxy delivers  the weakest constraints,  mainly due to its smaller sky area and number density. 

\begin{table}[!h] 
\centering 
\caption{Fractional errors $\sigma(\vartheta_\alpha)/\vartheta_\alpha$.} \label{tab3} 
\vspace*{0.2cm}
\begin{tabular}{|l|c|c|c|c|c|c|} 
\hline 
Sample~~&~~$\gamma$~~&~~$\sigma_{8,0}$~~&~~ $n_{s}$~~&~~$h$~~&~~$\Omega_{b0}$ & $\Omega_{c0}$ \\ \hline\hline
BGS   & 0.467  & 0.169 & 0.319 & 0.465 & 0.523 & 0.105 \\
ELG  & 0.119 & 0.018 & 0.066 & 0.102 & 0.107 & 0.020 \\ 
\hline\hline
HI gal.  & 0.664 & 0.25 & 0.921 & 1.445 & 1.586 & 0.239 \\ 
IM Band\,2. & 0.217 & 0.067 & 0.225 & 0.353  & 0.391 & 0.059 \\
IM Band\,1. & 0.061 & 0.013 & 0.050 & 0.077 & 0.084 & 0.018 \\  
\hline\hline  
BGS $\otimes$ HI gal.    & 0.228 & 0.096  & 0.295 & 0.465  & 0.520 & 0.089 \\
BGS $\otimes$ IM Band\,2 & 0.139 & 0.048  & 0.168  & 0.267 & 0.278 & 0.052 \\
ELG $\otimes$ IM Band\,1 & 0.047 & 0.008 & 0.037 & 0.057 & 0.062 & 0.013 \\
\hline \end{tabular}
\end{table}

When BGS is combined with HI IM Band\,2 (\autoref{fig5}) the situation changes. HI IM Band\,2 gives much better constraints than BGS, with an error on $\gamma$ less than half of the BGS error. This arises
despite the effects of foreground noise because HI IM Band\,2 covers a much larger area of the sky, which results in more Fourier modes that contribute to the Fisher analysis. In addition, foreground noise affects the largest scales where the $\gamma$ is not strong. When IM surveys are combined with spectroscopic galaxy surveys, the impact of foreground noise on the multi-tracer constraints is  further mitigated. \autoref{tab3} shows that the multi-tracer error on $\gamma$ is reduced by $\sim 40\%$ compared to the best single-tracer error from HI IM Band\,2. 

The best $\gamma$ precision is delivered at high redshifts by ELG\,$\otimes$\,IM\, Band\,1 (\autoref{fig4}).  The IM  Band\,1 error on $\gamma$ is $\sim 6\%$ while ELG produces about double this error. The multi-tracer reduces the error to $\sim 5\%$. 

\begin{table}[!ht] 
\centering 
\caption{Fractional errors on bias parameters.} \label{tab4} 
\vspace*{0.2cm}
\begin{tabular}{|l|l|c|c|} 
\hline 
Survey~~&~~Sample~~  & ~~${b_{A0}}$~~ &~~${b_{B0}}$\\ 
\hline\hline 
$g$~(DESI-like) 
                & BGS & 0.1719 & -  \\
                & ELG & 0.0169 & -  \\
 {$g$~(SKAO-like)} &HI gal. & 0.2691 & - \\  
                \hline \hline
$H$~(SKAO-like)  
                & IM Band\,2  & 0.0696 & - \\ & IM Band\,1  & 0.0138 & -\\ \hline \hline
{$g\otimes g'$} & BGS $\otimes$ HI gal.   & 0.0997  &  0.0989  \\
$g\otimes H$ & BGS $\otimes$ IM Band\,2 & 0.0501  &  0.0496  \\
& ELG $\otimes$ IM Band\,1 & 0.0083  &  0.0085 \\                            
\hline
\end{tabular}
\end{table}

\newpage

\newpage

\autoref{tab4} displays the fractional errors for the 2 bias nuisance parameters for the 3 survey combinations. The multi-tracer constraints on bias nuisance parameters are much tighter than those obtained from the individual single-tracers (compare \cite{Fonseca:2019qek,Viljoen:2020efi}).

\begin{figure}[! ht]
\centering
\includegraphics[width=7.5cm]{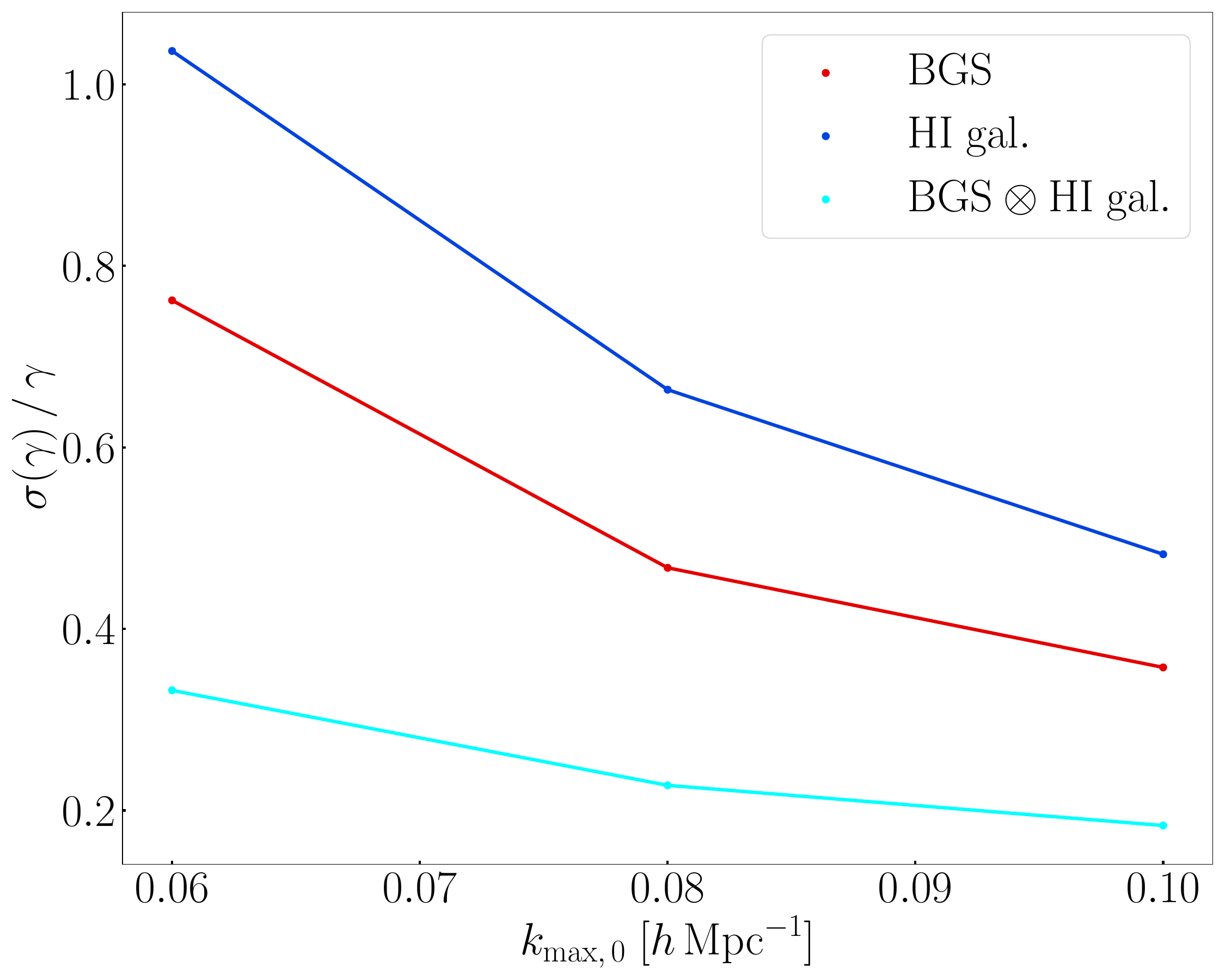}
\includegraphics[width=7.5cm]{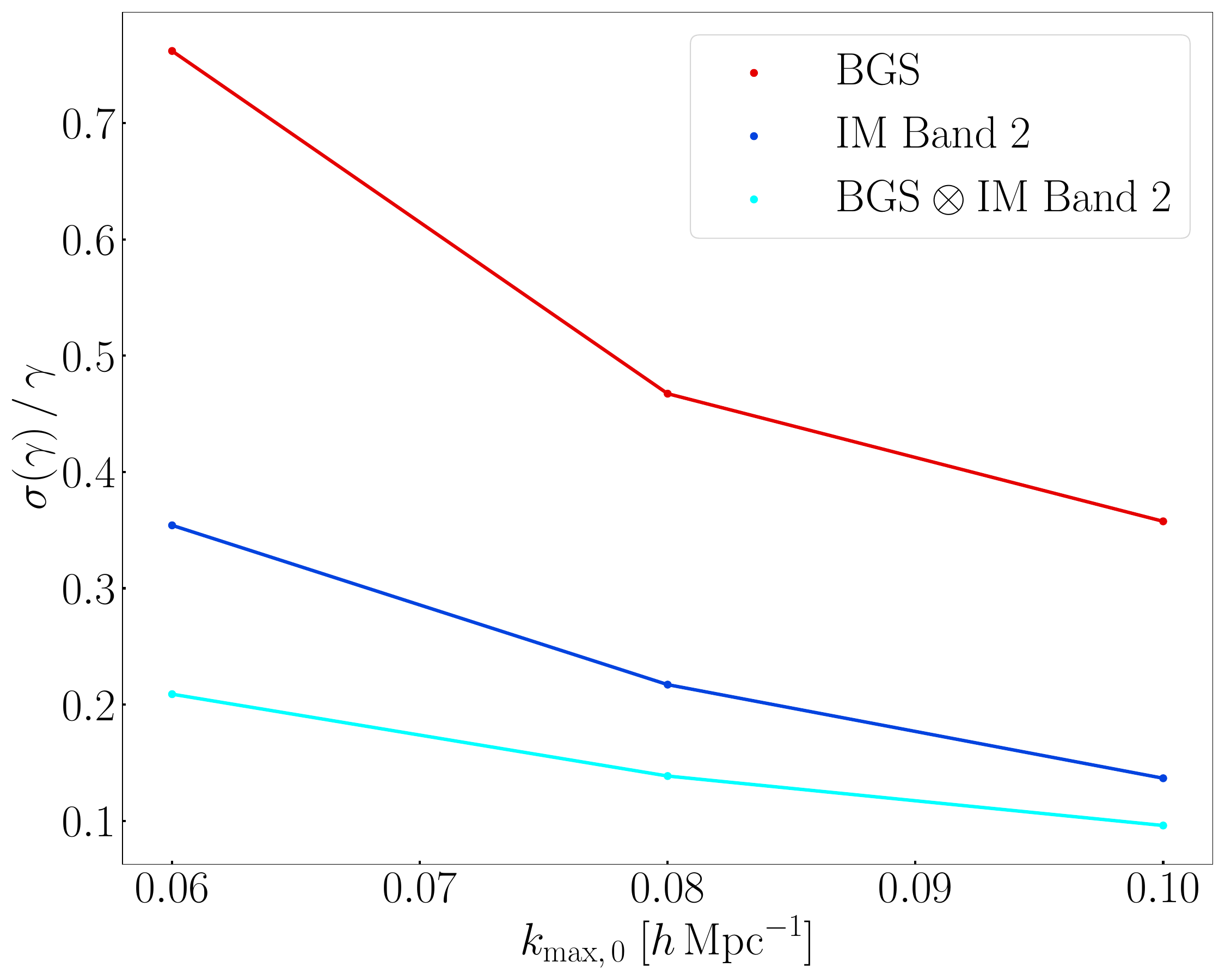} 
\includegraphics[width=7.5cm]{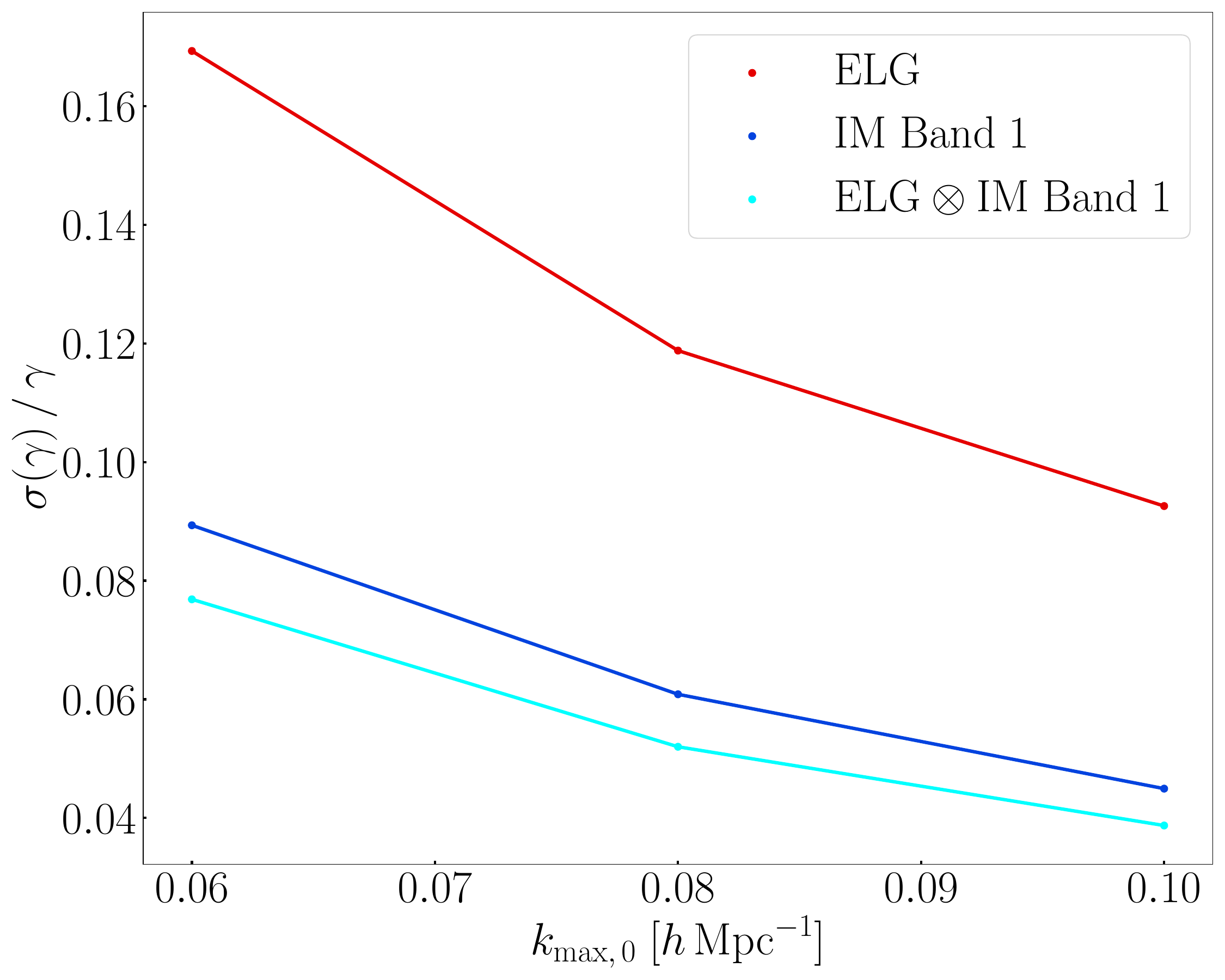} 
\bigskip
\caption{Change in fractional errors on $\gamma$ as a function of $k_{\rm max,0}$, for  the BGS and HI galaxy  ({top left}), BGS and HI IM Band\,2  ({top right}) and  ELG and HI IM Band\,1 ({bottom}) samples. Our choice is $k_{\rm max,0}=0.08\,h$/Mpc.} 
\label{fig6}
\end{figure}

All of our constraints are obtained from scales $k<k_{\rm max}$ where linear perturbations are accurate. 
In \autoref{fig6} we investigate the effect on the marginalised fractional error for $\gamma$ of changing $k_{\rm max,0}$ from its value given in \eqref{eq4.8}. The plots confirm that constraints are  sensitive to $k_{\mathrm{max,0}}$. We would unnecessarily lose information by reducing our  $k_{\rm max,0}$ value. On the other hand, increasing it leads to higher precision -- but at the risk of moving into the regime of nonlinear effects -- especially in RSD -- which requires much more effort to model. The multi-tracer has the advantage of allowing us to avoid these difficulties while at the same time delivering constraints that would not be possible with single tracers in the linear regime.

\newpage
\section{Conclusion}

{Using a simplified Fisher analysis we have estimated the multi-tracer constraints on the growth rate of large-scale structure, for pairs of tracer samples that are similar to those expected from the specifications of SKAO and DESI surveys. The  multi-tracer is known to be more effective, the more different are the pairs of surveys -- and this motivates our choice of DESI-like and SKAO-like samples. 

We applied a foreground-avoidance filter to the HI intensity mapping samples and included the effects of the radio telescope  beam, but we have not dealt with the many other systematics. Our aim is not a realistic forecast for specific surveys, but rather a proof of principle analysis to answer the question: {\em what is the potential of the multi-tracer to improve constraints using only linear scales?} 

By confining the signal to linear scales, we avoid the highly complex modelling, especially for RSD, that is required to access nonlinear scales. The cross-power spectra represent an additional complexity in the nonlinear regime \cite{Zhao:2020tis}.
Our Fisher analysis suggests that significant improvements in precision on the growth rate could be achieved by multi-tracing next-generation radio-optical pairs of samples. The details are summarised qualitatively in 
 \autoref{fig3}, \autoref{fig5} and \autoref{fig4}, and quantitatively in \autoref{tab3}.
 The biggest improvements, $\sim40-60\%$, are for the low-redshift pairs. 

This indicates that it is worthwhile to perform a more realistic analysis and derive more realistic forecasts. We leave this for further work.
Finally, we note that multi-tracer improvements can be delivered without requiring additional observational resources. }

\[\]
{\bf Acknowledgements}
We are supported by the South African Radio Astronomy Observatory (SARAO) and the National Research Foundation (Grant
No. 75415).

\clearpage
\bibliographystyle{JHEP}
\bibliography{reference_library}

\end{document}